\documentclass[aps,twocolumn]{revtex4}
\usepackage{epsfig}
\newcommand{\be}{\begin{equation}}
\newcommand{\ee}{\end{equation}}
\newcommand{\ba}{\begin{eqnarray}}
\newcommand{\ea}{\end{eqnarray}}

\def\mtil{m_{\tilde{\tau}}}

\def\msnu{m_{\tilde{\nu}}}

\begin{document}

\title{Weak Interactions of Supersymmetric Staus at High Energies}
\author{Yiwen Huang and Mary Hall Reno}
\affiliation{%
Department of Physics and Astronomy, University of Iowa,
Iowa City, Iowa 52242 USA}
\author{Ina Sarcevic and Jessica Uscinski}
\affiliation{%
Department of Physics, University of Arizona, Tucson, Arizona 85721,
USA}
\begin{abstract}

Neutrino telescopes may have the potential to detect the quasi-stable
staus predicted by some supersymmetric models. Detection depends
on stau electromagnetic energy loss and weak interactions.
We present results for the weak interactions contribution to the energy loss of
high energy staus as they pass through rock. We show that the neutral current weak interaction contribution is much smaller than photonuclear energy loss, however, the 
charged current contribution may become dominant 
process above an energy of $\sim 10^9$ GeV.
As a consequence, the stau range may be reduced above
$\sim 10^9$ GeV as compared to the range neglecting weak interactions.
We contrast this with the case of tau range, which
is barely changed with the inclusion of charged current
interactions.

\end{abstract}

\maketitle

\section{INTRODUCTION}

Interactions of very high energy neutrinos with
nucleons as they traverse the Earth are ideal probes of
physics beyond the Standard Model \cite{ringwald}.
High energy
neutrinos originate in interactions of
high energy cosmic rays with microwave background
photons (cosmogenic neutrinos), 
or they might be produced in astrophysical
sources such as active galactic
nuclei and gamma ray bursts \cite{learned}.
These neutrinos do not interact on their way to the Earth, and they arrive
undeflected by magnetic fields.  Once they reach the Earth,
they interact with nucleons in the Earth, or possibly
in the atmosphere.  In some supersymmetric models, neutrino 
interactions in Earth could produce heavy supersymmetric particles that
decay into quasi-stable sleptons. 

In the weak scale supersymmetric
models with the supersymmetry breaking scale larger than
$5 \times 10^6$ GeV, the next-to-lightest particle (NLP) is a charged slepton
(stau) which eventually
decays into the stable lightest supersymmetric particle (LSP),
the gravitino \cite{susy}.  
Due to very long lifetimes, staus 
may travel thousands of kilometers
through the Earth without decaying. Recently it was proposed that a
direct way of probing the SUSY breaking
scale in weak scale supersymmetry models would be to detect pairs of
charged tracks in neutrino detectors, such as IceCube, from
staus resulting from neutrino-nucleon interactions 
producing heavier supersymmetric particles \cite{chacko}.
This has been further explored in Refs. \cite{chacko2} and \cite{akr}.

Because of the small cross section, the
production of staus from downward neutrinos is negligible in comparison
with the background from the standard model processes.  However, upward
neutrinos producing staus could potentially be detectable because of the
effective detector volume that is enlarged
 by the long range of the stau.

Detection of staus produced in
 neutrino-nucleon interactions in Earth
depends strongly on the stau lifetime and range.
Thus it is crucial to determine
the energy loss and effective range of the high energy stau as it
traverses the Earth.
The average
electromagnetic energy loss of a particle which traverses a distance
$X$  is given by
\begin{equation}
-{dE\over dX} \simeq \alpha +\beta E
\end{equation}
where $E$ is the lepton energy, $\alpha$ represents the
ionization energy loss,
and $\beta$ is the radiative energy loss. To first approximation,
$\beta$ scales inversely with stau mass. 

In Ref. \cite{rss}, we
evaluated the electromagnetic energy loss of scalar
leptons more quantitatively. We showed that the photonuclear
interaction gives the largest contribution to $\beta$ for
stau energies between $10^6-10^{12}$ GeV. The range determined
by electromagnetic interactions is of order $10^4$ km.w.e. for
stau masses of a few hundred GeV. 
Interaction lengths from charged current weak interactions are
of the same order of magnitude.
In this paper, we evaluate the energy and mass dependence of the
stau weak interaction energy loss and attenuation, and we
show its relevance to the stau range.
 
Weak interaction cross sections, because of the massive vector
boson propagators, have a different mass dependence than the
electromagnetic energy loss parameter $\beta$. Depending on the
details of the supersymmetric couplings, we find
that weak effects may dominate
the stau range.
 
In Sec. II, we present results for neutral and charged current cross sections
and energy loss
for staus and discuss its mass dependence.  We make comparisons
with the lepton case.
We show the tau and stau range 
including weak interaction processes using a
one-dimensional Monte Carlo evaluation described in Ref. \cite{drss}
in Sec. III. 
The weak interaction effects in $\tau$ propagation
through the Earth are determined to be 
negligible for the energies considered here --
up to $10^{12}$ GeV. Our conclusions for the tau range differ from
estimates based on characteristic distance scales for the tau
\cite{fargion}.
We discuss implications of the stau range including maximal
charged current interactions
for IceCube and higher energy measurements such as by
ANITA in Sec. IV.

\section{Stau Energy Loss}
\label{sec:energyloss}

The energy loss parameter $\beta$ has contributions from
a variety of processes:
\begin{equation}
\label{eq:beta}
\beta^i (E) = {N_A\over A}\int_{y_{min}}^{y_{max}}
dy\  y{d\sigma^i(y,E)\over dy},
\end{equation}
where $y$ is the fraction of lepton energy loss in the radiative interaction,
\begin{equation}
y={E-E'\over E},
\end{equation}
for final stau energy $E'$. The superscript $i$ denotes
bremsstrahlung (brem) \cite{brem,brem2}, pair production (pair) \cite{pair}, 
photonuclear (nuc) \cite{drss, bugaev}
and weak (NC) processes for interactions of the initial particle
with a target nucleus. Avogadro's number is $N_A$ and the
atomic mass number of the target
nucleus is $A$.

At low energies,
where $\beta E\ll \alpha$, either the lifetime
or ionization energy loss determines the stau range,
which scales linearly with energy \cite{drss}.
The
ionization
energy loss parameter $\alpha$ is nearly constant as a function of
mass of the particle, namely \cite{ionization}
\begin{equation}
\alpha\simeq 2\times 10^{-3} \ {\rm GeV cm^2/g}\ .
\end{equation}
At energies above $10^6$ GeV, energy loss of leptons and staus
is dominated by the electromagnetic radiative processes.
For staus, weak interactions may also become important. We review
next the weak interaction cross sections for scalars.

\subsection{Weak Interaction Cross Sections}

Neutral current and charged current cross sections are relevant
in two different ways. Neutral current interactions only shift
the stau energy. The neutral current interactions can be incorporated
into $\beta$. By contrast, the charged current interactions remove
the stau, in the process producing a sneutrino. The sneutrino
then decays, presumably to another stau. We do not include 
stau regeneration because of the several steps decreasing energy.
Neutral current and charged current cross sections of staus
are shown below, as well as the results for the cross sections for taus.

\subsubsection{Neutral Current Cross Sections}

The neutral current cross section that
describes the interactions of charged
sleptons ($\tilde{\tau}$) with nucleons via exchange of $Z^0$ boson
is given by 
\begin{eqnarray}
\frac{d^2\sigma ^{NC} (\tilde{\tau}N)}{dxdy} &=&
\frac{G_F^2}{\pi} ME \Bigl(\frac{M_Z^2}{Q^2+M_Z^2}\Bigr)^2
\sin^2{2\theta_W} \nonumber \\ &\cdot & \Bigl(\alpha_f
+\beta_f\cos2{\theta_f}
\Bigr)^2 \Bigl[ 2 x\Bigl(1-\frac{y}{2}\Bigr)^2  \nonumber\\
&-&\Bigl(\frac{xy^2}{2}+\frac{m^2_{\tilde{\tau}}y}{ME}\Bigr) \Bigr]F_1^{NC}\ .
\end{eqnarray}
The parton fractional momentum is $x$ and $y$ is the fraction
of slepton energy loss.
The quantities $\alpha_f$ and $\beta_f$ are the couplings of staus to 
gauge bosons \cite{howie}
\begin{eqnarray}
\alpha_f=\frac{1}{4}\Bigl(3\tan\theta_W-\cot\theta_W\Bigr)
\end{eqnarray}
 and
\begin{eqnarray}
\beta_f=\frac{1}{4}\Bigl(\tan\theta_W+\cot\theta_W\Bigr)\ .
\end{eqnarray}
The scalar partner of the right-handed tau may not be a mass eigenstate.
The angle $\theta_f$ parameterizes the mixing between scalar partners
of the right-handed and left-handed tau, where $\sin\theta_f=0$
means that the mass eigenstate quasi-stable stau is purely 
made of the partner of the right-handed tau. In principle, $\sin\theta
_f$ need not equal zero. We take
$\sin\theta_f=1$ for the neutral-current process in the figures
below to evaluate the maximal effect in the charged current case,
since $W$'s couple only to left-handed fermions and their scalar
partners. 
The range of $(\alpha_f+\beta_f\cos2{\theta_f})^2$ is such that
\begin{equation}
0\leq \Big[(\alpha_f+\beta_f\cos 2\theta_f)^2/(\alpha_f+\beta_f)^2
\equiv r_{NC}\Bigr] \leq
1.38\ .
\end{equation}

For the neutral current structure functions, 
we have taken $2xF_1=F_2$ here and for charged current interactions.
For neutral currents,
\begin{eqnarray}
\label{eq:f1}
F_1^{NC} &=& \frac{1}{2}(v_i^2+a_i^2)[q_i(x,Q^2)+\bar{q}_i(x,Q^2)]
\end{eqnarray}
with 
\begin{eqnarray}
v_i &=& T_3 - 2 e_i\sin^2\theta_W \\
a_i &=& T_3
\end{eqnarray}
for weak isospin assignments $T_3=\pm 1/2$ and electric charge
$e_i$.
We use CTEQ6 parton distribution functions \cite{cteq6}
with a power law extrapolation of these distributions for $x < 10^{-6}$
of the form $x^{-\lambda_i}$, where $i$ denotes quark or
antiquark flavor
\cite{gqrs,reno}.  The values
we use for $\lambda_i$ are given by:
$$
\begin{array}{lrrr}
\ \ & u\bar{u},d\bar{d} \quad \quad \quad &
s\bar{s},b\bar{b} \quad \ & c\bar{c}\quad \quad \\
\lambda_i \ \ & -0.0276\cdot\ln{Q}+0.1784 & \ \lambda_u + 0.0054
& \lambda_u+0.0094.\\
\end{array}
$$
The kinematic limits on the variables of
integration,
$y$ and $Q^2$ (for small $y$) are given by
\begin{eqnarray}
\frac{\mtil^2y^2}{1-y} \ \leq \ Q^2  &\leq&
4E^2(1-y) - \frac{\mtil^2(2-y)^2}{1-y}
\nonumber\\\frac{\Bigl((M+m_\pi)^2-M^2\Bigr)}{2ME}\ &\leq& \ y \leq 1-
\frac{\mtil}{E}
\label{eq:nclimits}
\end{eqnarray}

In Fig. 1 we show the neutral-current cross sections for stau masses of $50$ GeV and $250$ GeV,
for $\sin\theta_f=1$. We also show
the muon, tau and neutrino neutral current
cross sections for comparison.  
We note that the cross section for
taus is almost indistiguishable from the muons
because the masses of taus and muons are
small compared to the energy considered.  The 
stau cross sections have weak $m_{\tilde{\tau}}$ dependence.

We see that the stau NC cross section is almost an order of magnitude
smaller than the neutrino case for $\sin\theta_f=1$
and $m_{\tilde{\tau}}=250$ GeV at
$E=10^6$ GeV.  The difference comes from the
couplings as well as the $y$ dependence of the differential cross section.
The ratio of neutrino to stau neutral-current couplings, including spin
averaging, yields about a factor of 2.   At small
$y$, the differential cross section, $d\sigma/dy$, for neutrinos is larger by about a factor 
of 5 than for the stau.
The average $y$ for staus decreases with increasing stau mass from
$<y> \approx 0.13$ for $\mtil$ = 50 GeV to $<y> \approx 0.06$ for $\mtil$ =
250 GeV.  
\begin{figure}[h]
\begin{center}
\epsfig{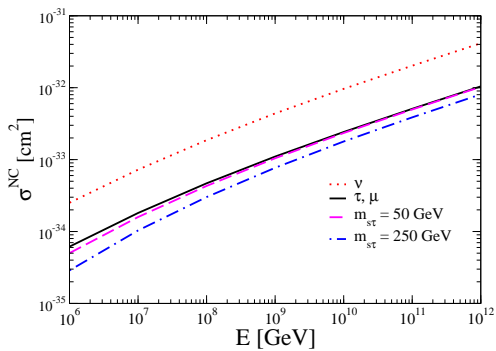}
\end{center}
\caption{Neutral-current cross sections for neutrino, tau, muon,
and stau for $\mtil=50$ GeV and $\mtil=250$ GeV, with $\sin\theta_f=1$.}
\end{figure}

\subsubsection{Charged Current Cross Sections}
The charged-current cross section of staus with
nucleons via exchange of charged boson is given by
\begin{eqnarray}
\frac{d^2\sigma ^{CC} (\tilde{\tau}N)}{dxdy} &=&
\frac{G_F^2}{\pi}\Bigl(\frac{M_W^2}{Q^2+M_W^2}\Bigr)^2
{\sin^2{\theta_f}}ME\nonumber\\
&\cdot&F_1^{CC} \Bigl[2 x \Bigl(1-\frac{y}{2}\Bigr)^2 \nonumber\\&-&
\frac{y}{2ME}\Bigl(m^2_{\tilde{\tau}}+\msnu^2+MExy\Bigr)\Bigr].
\end{eqnarray}
Since we are interested
in considering the upper limit on the stau cross section, we take
$\sin^2{\theta_f}=1$, however, the value of $\sin\theta_f$ is unknown.
For the charged current, the conventional
normalization of the structure functions is
\begin{eqnarray}
F_1^{CC} &=& [q_i(x,Q^2)+\bar{q}_j(x,Q^2)]\nonumber\\
F_3^{CC} &=& 2[q_i(x,Q^2)-\bar{q}_j(x,Q^2)]
\end{eqnarray}
and we use the kinematic limits
\begin{eqnarray}
\frac{\msnu^2y}{1-y} - \mtil y \ \leq \ Q^2  &\leq&
4E^2(1-y) - m^2_{\tilde{\tau}}
(2-y)\nonumber \\&-& \frac{\msnu^2(2-y)}{1-y}
\nonumber\\\frac{\Bigl((M+m_\pi)^2-M^2\Bigr)}{2ME}\ \leq \ y &\leq& 1-
\frac{\msnu}{E},
\end{eqnarray}
where $\mtil$ represents the mass of the incoming stau and
$\msnu$ represents the outgoing sneutrino.
For the stau process we take 
$\msnu - \mtil = 50$ GeV, with $\mtil$ as $50$ GeV and $250$ GeV.

In Fig. 2 we show the charged-current (CC) cross
sections for the stau with mass $50$ GeV and $250$ GeV,
and for the tau, muon and neutrino for comparison.  
We note again that the cross section for stau
has a weak mass dependence.  The cross sections for taus
and muons are indistinguishable due to the small masses relative to the
energies we consider.  The charged 
lepton CC cross section is about a factor of 2 smaller
than the neutrino case, due to the spin averaging.
The energy dependence of the stau cross section is stronger than for 
the tau 
and muon.

\begin{figure}[h]
\begin{center}
\epsfig{file=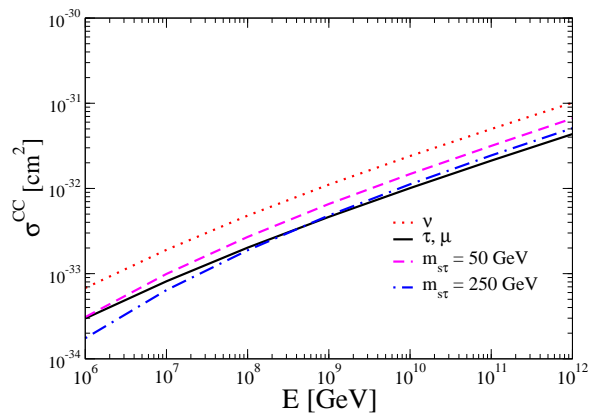,width=2.55in,angle=270}
\end{center}
\caption{Charged-current cross sections for neutrino, tau, muon, 
and stau for $\mtil=50$ GeV and $\mtil=250$ GeV, with $\sin\theta_f=1$.}
\end{figure}

\subsection{Application to Energy Loss}

\begin{figure}[h]
\begin{center}
\epsfig{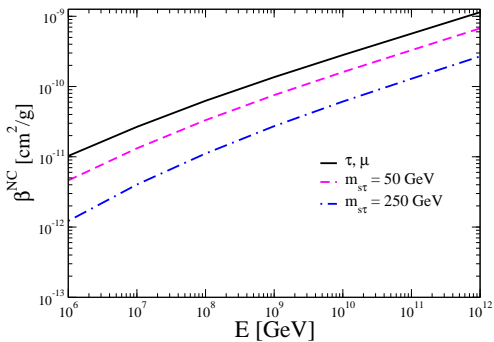}
\end{center}
\caption{Neutral-current $\beta$ for tau, muon,
 and stau for $m_{\tilde\tau}=50$ GeV and $m_{\tilde\tau}=250$ GeV, with $\sin\theta_f=1$.}
\end{figure}

The energy loss $\beta$ for neutral current interactions
can be found at a fixed initial charged lepton
or stau energy from Eq. (\ref{eq:beta}).
In Fig. 3 we show our results for $\beta^{NC}$ for muon,
tau and stau with masses $50$ GeV and $250$ GeV, with $\sin\theta_f=1$.  
The values of $\beta^{NC}$ for the muon and tau are very close 
in value, as they were for the cross sections.  Using the relation
$$\beta\simeq N\langle y\rangle \sigma(E)\ ,$$ we estimate the average value
for $y$ to be about $0.2$ for tau and muon neutral-current interaction.
We see that the stau case is smaller
than tau by about a factor of 3 at $10^6$ GeV when we use a stau mass of 
$50$ GeV. We show below that the photonuclear contribution to $\beta$ is
at least an order of magnitude larger than $\beta^{NC}$ for staus.
For muons and taus, the photonuclear $\beta^{nuc}$ is larger 
than $\beta^{NC}$ by a factor
$\sim 10^3-10^4$.

In Fig. 4 we show the mass dependence of $\beta^{NC}$ for different 
initial energies.
We note that the mass dependence is weaker than ${1}/{\mtil}$ for
$\mtil\leq 200$ GeV. For masses larger than 
200 GeV, $1/\mtil$ scaling works reasonably well.

\vskip 0.1in
\begin{figure}[h]
\begin{center}
\epsfig{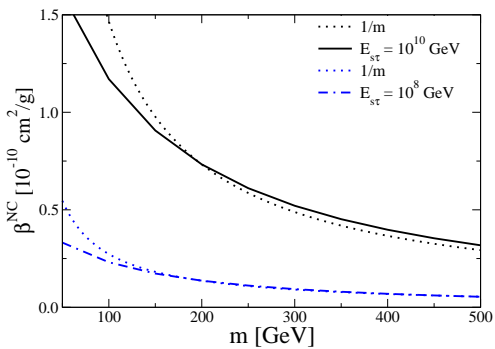}
\end{center}
\caption{Mass dependence of neutral-current $\beta$ for stau with a
stau mass range of $m_{\tilde\tau}=50$ GeV to $m_{\tilde\tau}=500$ GeV, with 
$\sin\theta_f=1$.}
\end{figure}


In charged-current processes the initial and final states are 
different so we use
$N\sigma^{CC}$, the inverse of the
effective interaction length, instead of $\beta$. 
We show in Fig. 5 the results for $N\sigma^{CC}$
together with $\beta^{nuc}$.  These figures are plotted using
$\sin\theta_f=1$ and for a sneutrino mass 50 GeV more massive
than the stau. For the sneutrino mass between
5 GeV and 150 GeV more than the stau mass, the cross section
changes by a factor of $\sim 1/2-2$. In the following, we show
only a 50 GeV mass difference.

Comparing the scales associated
with weak interaction and
electromagnetic 
processes shows that the charged-current interactions become significant
at higher energies, in contrast to the neutral current
interactions.  For a stau mass of $250$ GeV and $\sin\theta_f=1$, this corresponds to 
energies higher than about $4\times 10^9$ GeV.  For the lighter $50$ GeV 
stau mass, the charged current process does not contribute significantly for 
energies up to $10^{12}$ GeV. The opportunity for charged current interactions to become
relevant comes from the fact that the charged current cross section is less sensitive
to the stau mass than the photonuclear energy loss parameter $\beta$.

A comparison of Figs. 5 and 3 verifies our assertion that the weak neutral current
contribution to $\beta$ for staus is not important. This is also true for tau energy loss.

\begin{figure}[h]
\begin{center}
\epsfig{file=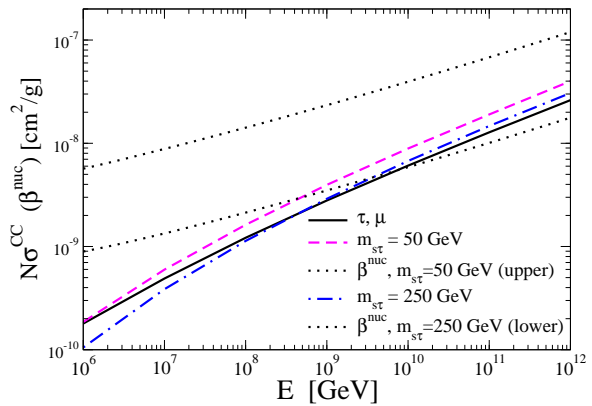,width=2.55in,angle=270}
\end{center}
\caption{Charged-current interaction length for tau, muon, 
and stau for $m_{\tilde\tau}=50$ GeV and $m_{\tilde\tau}=250$ GeV, with $\sin\theta_f=1$.
The sneutrino mass is $\mtil+50$ GeV.
Also shown are photonuclear energy loss parameters for the two stau masses.}
\end{figure}

\section{Stau and Tau Range}

\begin{figure}[h]
\begin{center}
\epsfig{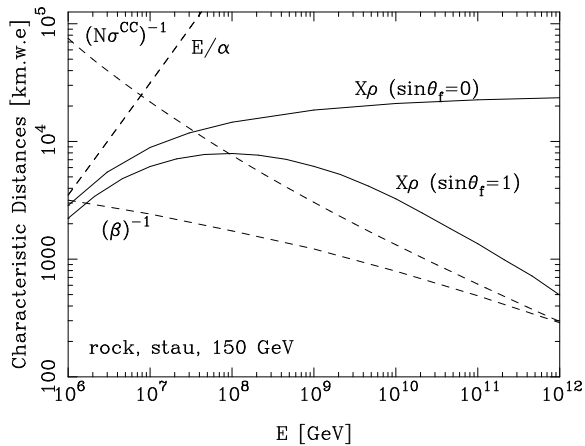}
\end{center}
\label{fig:s150dist}
\caption{Characteristic
distances in kilometers water equivalent units (dashed)
and ranges (solid)
for  staus in rock, for $\mtil=150$ GeV, 
$\sin\theta_f=0$ and 1, and $\sqrt{F}=10^7$ GeV. The minimum stau
energy is $E_0=10^3$ GeV. The sneutrino mass is $\mtil+50$ GeV.}
\end{figure}

Because $\beta^{NC}\ll \beta^{nuc}$ for staus, taus and muons, we
neglect neutral current interactions in our evaluation of the
range. For staus, the charged current interaction length is roughly
comparable to $1/\beta$ for some stau masses, so we include
the charged current interactions in our Monte Carlo evaluation
of the particle range. Details of the Monte Carlo evaluation
appear in Refs. \cite{drss} and \cite{rss}. The Monte Carlo
computer program computes survival probabilities  $P(E,E_0,X')$
for an
particle incident with energy $E$ which survives a distance
$X'$ with $E>E_0$. The range is defined by
\begin{equation}
X(E,E_0)\equiv \int dX' P(E,E_0,X')\ .
\end{equation}
We have taken $E_0=10^3$ GeV for the figures shown here for staus.

In Fig. 6, we show characteristic distances
associated with stau interactions in rock. The curves are evaluated
for $\mtil=150$ GeV and the decay parameter $F^{1/2}=10^7$ GeV,
where the lifetime is determined by
\begin{equation}
c\tau = \Biggl(\frac{F}{10^{14}\ {\rm GeV}^2}\Biggr)^2 \Biggl(\frac{100\ {\rm GeV}}{\mtil}\Biggr)^5
\, 10\ {\rm km}\ .
\end{equation}
The lifetime is not relevant for this energy range for $\mtil=150$ GeV
and $F^{1/2}=10^7$ GeV:
$E c\tau\rho/\mtil  \simeq 3\cdot 10^4$ kmwe for $E=10^6$ GeV. A distance
which is relevant, which also grows with energy, depends on the ionization
energy loss parameter $\alpha$ through $d\sim E/\alpha$, shown in the
figure. We also show the charged current interaction length $(N\sigma^{CC})^{-1}$ and the distance characterized by $\beta^{-1}$,
the electromagnetic energy loss parameter.  We show the range for
150 GeV staus with no charged current contributions ($\sin\theta_f=0$)
and with maximal charged current contributions ($\sin\theta_f=1$).
At low energies, the ionization energy loss dominates, but for $E\sim
10^8$ GeV, the charged current interaction dominates the
evaluation of the range if $\sin\theta_f=1$. The range does
not precisely equal the charged current interaction length because
the electromagnetic energy loss is still a factor, shifting the
initial stau energy to lower energies.

Fig. 7 shows the stau ranges in
rock for $\mtil=150$ and 250 GeV, again
for minimum stau energy of $10^3$ GeV and $F^{1/2}=10^7$ GeV.
The upper curves show the range when charged current interactions
are vanishing, while the lower curves have maximal charged current
interactions. 

Finally, in Fig. 8 we show the characteristic distances for
taus in rock and the tau range. 
The lifetime governs the range at low energies, while
electromagnetic energy loss dominates at high energies. Because the 
charged current interaction length is small compared to $(\beta\rho)
^{-1}$, the tau range
changes very little with charged current interactions included.
We do not find a decrease in the range 
in rock or water near 
$E=10^{12}$ GeV as suggested in Ref. \cite{fargion}. 
This is due to the fact that using just the scales, e.g.,
$(\beta\rho )^{-1}$ or the CC interaction length, is insufficient
to accurately compute the range. We have directly evaluated
the tau range via Eq. (16), where the probability 
includes stochastic effects in the tau propagation.

\begin{figure}[h]
\begin{center}
\epsfig{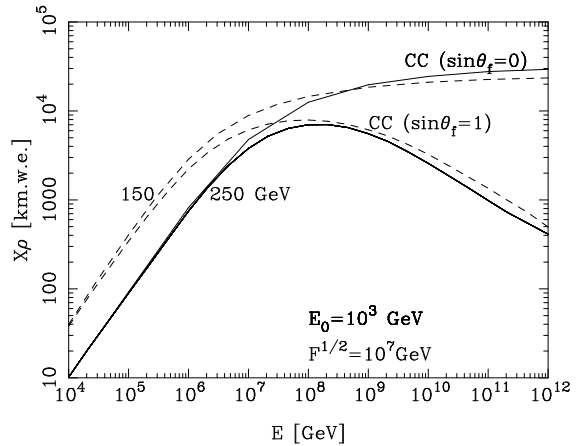}
\end{center}
\caption{Range of stau in rock, for $\mtil=150$ and 250 GeV, $\sin\theta_f=0$ and 1.
The lifetime is here governed by $\sqrt{F}=10^7$ GeV, and the minimum stau
energy is $E_0=10^3$ GeV. The sneutrino mass is $\mtil+50$ GeV.}
\end{figure}

\begin{figure}[h]
\begin{center}
\epsfig{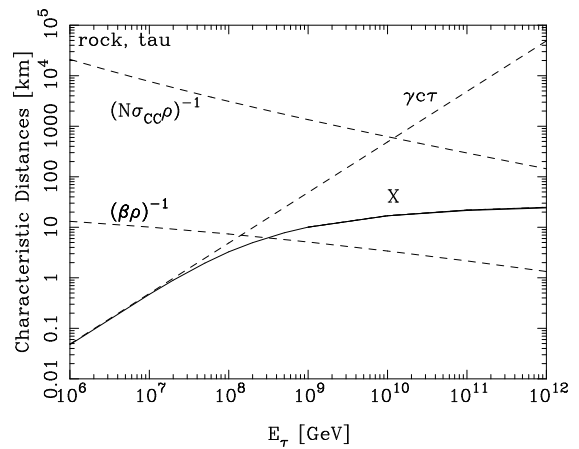}
\end{center}
\caption{Characteristic
distances (dashed) and tau range in rock (solid) in km. 
By including the charged current interaction
length, the tau range is unchanged on the scale of the figure.}
\end{figure}

\section{Conclusions}

We have shown that weak interactions have the potential to play
an important role in stau detection by neutrino telescopes, however,
the effect is strongly energy dependent.
Our results are based on a sneutrino mass 
50 GeV more massive than the stau, and we considered $\sin\theta_f=1$.
 
Recent
work on stau signals \cite{chacko2,akr} 
has focused on the IceCube detector. The stau
pair event rate depends on the effective volume for creation
of a pair of staus. The effective volume scales approximately
with the stau range which depends on energy.

Refs. \cite{chacko2,akr} show that the
signal events come from energies fairly near the threshold for squark production
because of the falling neutrino fluxes. In these studies, an incident
$E^{-2}$ neutrino flux is assumed. The threshold depends on squark masses:
for $m_{\tilde{q}}=300-900$ GeV, the energy threshold is between a few$\times
10^5-10^6$ GeV. For $E=10^6$ GeV, even a maximal charged current cross sections
with $m_{\tilde{\nu}}=\mtil +50$ GeV doesn't affect the stau energy range dramatically. For $\mtil=150$ GeV, maximal CC interactions reduce
the range by 22\% at $10^6$ GeV, and by 31\%
at $10^7$ GeV.  
The effect is less pronounced for $\mtil=250$ GeV. 
When $\sin \theta_f=1$ for
$\mtil=250$ GeV and $F^{1/2}=10^7$ GeV, the range is reduced by
9\% and 21\% for $E=10^6$ and $10^7$ GeV, respectively. 
Based on these reductions in the range, the thresholds in
the $10^5-10^6$ GeV energy range and steeply falling fluxes, 
event rate estimates without including weak interactions are
reasonably reliable.

Efforts to try to detect staus with higher energy thresholds are potentially
strongly influenced by charged current interactions, where the range can be as much
as two orders of magnitude shorter than the range evaluated without charged
current interactions. A detector such as the Antarctic
Impulse Transient Array (ANITA) \cite{anita} is sensitive to
stau energies larger than $\sim 10^8$ GeV. Designed to use
a radio antenna suspended by a balloon $\sim 37$ km over the
south pole ice, the primary goal of the experiment is to detect
cosmogenic neutrinos incident just below the horizon which interact
with the ice. The goal is to detect 
the radio Cherenkov signal produced by neutrino weak
interactions which refracts 
on its way out of the ice. Staus would also make a signal by
weakly interacting in the ice.

Weak interactions play a role for ANITA signals in two ways. 
As noted above, the most important feature of weak interactions
is to produce the signal itself.  
For electromagnetic interactions, 
only a small amount of energy will be deposited in the shower over the area
of $\sim 10^6$ km$^2$.  Neutral current interactions 
have a larger 
energy deposition necessary for detection. Charged current interactions
have the largest energy deposition. 
Only neutral current processes will contribute to the signal if 
$\sin\theta_f=0$, while charged current processes contribute with
increasing  values of
$\sin\theta_f$.  
The maximum contribution from charged current interactions occur
when 
$\sin\theta_f=1$.   For this maximum value of the mixing angle,
$\sigma^{CC}/\sigma^{NC}\sim 10$, meaning that the probability for interactions
of staus in the ice is increased by a factor of 10 over the case of 
$\sin\theta_f=0$.

A second effect for ANITA signals is that
charged current weak interactions 
may attenuate the stau flux in transit to the
ice in view of the detector.  By incorporating the attenuation of 
the GZK neutrino flux \cite{gzk} as it traverses the Earth, 
the stau production cross section \cite{chacko} 
and the stau interactions on 
the way to the detector, we find that the stau flux attenuation 
at $10^9$
GeV ranges from 1 at $0^\circ$ to $\sim 1/3$ at a $10^\circ$ angle measured
relative to the horizon \cite{future}.  
For energies of 
$10^{10}$ GeV, attenuation causes the stau flux to be lowered 
by a factor of $\sim 1/10$ 
with maximal charged current interactions
for an angle of 
$10^\circ$ below the horizon.
The factor of 10 increase in the signal due to the stau CC interactions
in ice is sufficient to compensate for the attenuation as well as enhance the  signal when the full range of energy is considered.

To summarize, weak interaction effects are small in the range
of energies relevant to the IceCube detector. Recent event rate
estimates \cite{chacko,chacko2,akr} for IceCube will be reduced by less
than $\sim 30\%$ for $E=10^6-10^7$ GeV 
by including maximal weak interactions.
The potential for observing staus at higher energies, for example,
by the ANITA detector, is enhanced by maximal weak interactions.
The enhancement is due to the high energy charged current cross 
section which becomes increasingly important as the stau energy
increases above $10^8$ GeV.
A detailed investigation of the ANITA signal from
stau pairs, and for the proposed ARIANNA \cite{arianna} telescope
is in progress \cite{future}.
Signals of staus from several energy regimes may impose constraints
on this class of supersymmetry models with quasi-stable staus
in the future.

\acknowledgments
We thank John Beacom for inspiring us to look at weak interaction
stau energy loss and for many useful discussions.
This work was supported in part by DOE contracts DE-FG02-91ER40664,
DE-FG02-04ER41319 and DE-FG02-04ER41298 (Task C).

\end{document}